\documentclass[10pt, conference, letterpaper]{IEEEtran}
\IEEEoverridecommandlockouts
\usepackage[nospace]{cite}
\usepackage{amsmath,amssymb,amsfonts}
\usepackage{mathtools}
\usepackage{graphicx}
\usepackage{textcomp}
\usepackage{multirow}
\usepackage{pifont}
\usepackage{xcolor}
\usepackage{diagbox}
\usepackage{tabularray}
\usepackage{url}

\usepackage{algorithm}
\usepackage[noend]{algpseudocode}
\usepackage{amssymb}
\usepackage{amsthm}

\usepackage{caption}
\usepackage{subcaption}
\usepackage[english]{babel}
\usepackage{amsthm}

\def\BibTeX{{\rm B\kern-.05em{\sc i\kern-.025em b}\kern-.08em
    T\kern-.1667em\lower.7ex\hbox{E}\kern-.125emX}}
\begin{document}

\title{Parm: Efficient Training of Large Sparsely-Activated Models with Dedicated Schedules 
}

\author{\IEEEauthorblockN{Xinglin Pan\IEEEauthorrefmark{2}\IEEEauthorrefmark{3}, Wenxiang Lin\IEEEauthorrefmark{4}, Shaohuai Shi\IEEEauthorrefmark{4}\IEEEauthorrefmark{1}, Xiaowen Chu\IEEEauthorrefmark{2}\IEEEauthorrefmark{5}\IEEEauthorrefmark{1}\thanks{*Corresponding author.}, Weinong Sun\IEEEauthorrefmark{5}, Bo Li\IEEEauthorrefmark{5}\\}
\IEEEauthorblockA{\IEEEauthorrefmark{2}
Data Science and Analytics Thrust, The Hong Kong University of Science and Technology (Guangzhou)\\
\IEEEauthorrefmark{3}Department of Computer Science, Hong Kong Baptist University\\
\IEEEauthorrefmark{4}School of Computer Science and Technology, Harbin Institute of Technology, Shenzhen\\
\IEEEauthorrefmark{5}Department of Computer Science and Engineering, The Hong Kong University of Science and Technology\\
xpan413@connect.hkust-gz.edu.cn, wenxianglineut@163.com, 
	shaohuais@hit.edu.cn,\\  xwchu@ust.hk, wnsun@ust.hk, bli@cse.ust.hk}
}

\maketitle

\begin{abstract}
Sparsely-activated Mixture-of-Expert (MoE) layers have found practical applications in enlarging the model size of large-scale foundation models, with only a sub-linear increase in computation demands. Despite the wide adoption of hybrid parallel paradigms like model parallelism, expert parallelism, and expert-sharding parallelism (i.e., MP+EP+ESP) to support MoE model training on GPU clusters, the training efficiency is hindered by communication costs introduced by these parallel paradigms. To address this limitation, we propose Parm, a system that accelerates MP+EP+ESP training by designing two dedicated schedules for placing communication tasks. The proposed schedules eliminate redundant computations and communications and enable overlaps between intra-node and inter-node communications, ultimately reducing the overall training time. As the two schedules are not mutually exclusive, we provide comprehensive theoretical analyses and derive an automatic and accurate solution to determine which schedule should be applied in different scenarios. Experimental results on an 8-GPU server and a 32-GPU cluster demonstrate that Parm outperforms the state-of-the-art MoE training system, DeepSpeed-MoE, achieving 1.13$\times$ to 5.77$\times$ speedup on 1296 manually configured MoE layers and approximately 3$\times$ improvement on two real-world MoE models based on BERT and GPT-2.
\end{abstract}

\section{Introduction}
The generalization performance of foundation models is predominantly influenced by their scale~\cite{DBLP:Kaplan2020Scaling} (e.g., GPT-2~\cite{radford2019language} with 1.5 billion parameters and GPT-3~\cite{DBLP:Brown2020GPT3} with 175 billion parameters). However, the immense size of these models inevitably demands substantial computing resources. For instance, training GPT-3 requires $3.14\times 10^{23}$ FLOPs~\cite{DBLP:Brown2020GPT3}, which is a daunting amount of computation. This has raised concerns about the feasibility of expanding these models further. In recent years, sparsely-activated Mixture-of-Experts~(MoE) layers have gained significant attention for increasing model size while demanding only a sub-linear increase in computation~\cite{DBLP:Chowdhery2022plam, DBLP:danny2023palme, DBLP:Shazeer2017outrageously, DBLP:Lepikhin2021gshard}. An MoE layer incorporates multiple experts, each typically a feed-forward block trained for specific subtasks. It employs a gating mechanism, often implemented using softmax activation~\cite{DBLP:Lepikhin2021gshard}, to dynamically determine which experts are trained with which samples. As a result, it enables scaling up the model size while requiring a relatively smaller computing cost. For example, Switch Transformer~\cite{DBLP:Fedus2022switch} scales to 1.5 trillion parameters with MoE.

\begin{figure}[!t]
	\centering
	\includegraphics[width=0.6\linewidth]{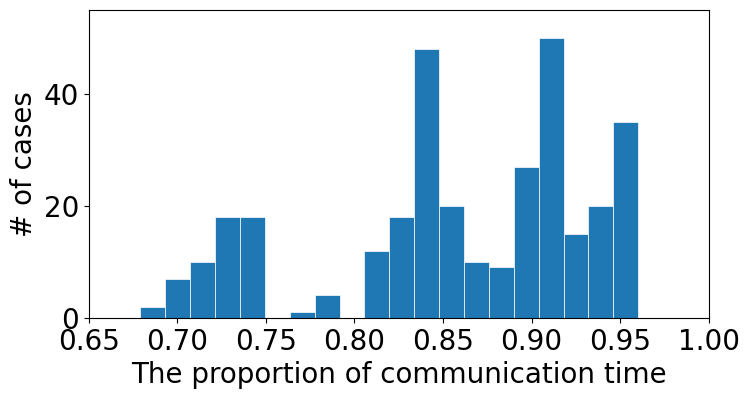}
	\caption{The communication time ratio varies across different configurations, ranging from 67.92\% to 96.02\% when using 32 Nvidia GeForce RTX 2080Ti GPUs. Detailed configurations for test cases can be found in Table~\ref{tab:moe-configs}. }
\vspace{-3mm}
\label{fig:communication_time_occupation}
\end{figure}

As the scale of foundation models continues to grow, various parallel paradigms are emerging to meet the growing computational demands. For traditional models, three commonly used parallel paradigms (also known as 3D parallelism~\cite{narayanan2021efficient}) are: \textit{Data Parallelism} (DP)~\cite{DBLP:dean2012large}, \textit{Model Parallelism} (MP)~\cite{DBLP:dean2012large}, and \textit{Pipeline Parallelism} (PP)~\cite{DBLP:huang2019gpipe}. For MoE-based models, two additional parallel paradigms are involved: \textit{Expert Parallelism} (EP)~\cite{DBLP:Shazeer2017outrageously} and \textit{Expert-Sharding Parallelism} (ESP)~\cite{DBLP:Rajbhandari22deepspeedmoe}. EP places experts across multiple devices when a single device cannot accommodate all experts. It dynamically assigns each input to particular experts using an AlltoAll collective communication known as \textit{AlltoAll Dispatch}. Another AlltoAll collective communication, named \textit{AlltoAll Combine}, then gathers the computed results from experts. On the other hand, ESP further divides a single expert into multiple shards and executes across multiple devices in parallel when a single expert cannot be stored on one device. 
EP and ESP are commonly used in conjunction to distribute the workload of an MoE layer across multiple devices, thereby enabling efficient parallel computing~\cite{DBLP:Rajbhandari22deepspeedmoe,hwang2023tutel} in large-scale training scenarios.

Nevertheless, the training performance of MoE-based models in GPU or TPU clusters is hampered by the communication time introduced by various parallel paradigms. As reported in~\cite{DBLP:Lepikhin2021gshard, hwang2023tutel, DBLP:liu2022gating}, the AlltoAll communication time occupies 30\% to 60\% of the overall time of the MoE layers on high-end GPU or TPU clusters. This situation worsens when using multiple parallel paradigms such as MP, EP, and ESP (MP+EP+ESP) simultaneously. To delve deeper into this issue, we measured the communication time (including communication introduced by DP, MP, EP, and ESP) for 324 different configurations of MoE layers using 32 Nvidia GeForce RTX 2080Ti GPUs, as shown in Fig.~\ref{fig:communication_time_occupation}. Our findings reveal that the communication time significantly dominates the training time of an MoE layer. In summary, training large MoE-based models on GPU clusters poses a significant performance challenge.

Prior research has endeavored to tackle the performance challenges in training MoE-based foundation models from three distinct directions. First, algorithm-level innovations, exemplified by works like \cite{DBLP:Lepikhin2021gshard} and \cite{DBLP:Fedus2022switch}, aim to design load-balancing gates that ensure more balanced network traffic across GPUs. Second, system-level studies, such as \cite{DBLP:ma2022bagualu} and \cite{DBLP:nie2022hetumoe}, focus on devising AlltoAll algorithms that are tailored to specific network topological architectures. Lastly, scheduling techniques \cite{DBLP:zhang2022accelerating,DBLP:Rajbhandari22deepspeedmoe,shi2023pipemoe,li2023lina} have been proposed to explore the overlap of AlltoAll communications and computing tasks in the MoE layer to enhance GPU utilization. However, these studies have primarily concentrated on mitigating AlltoAll communication costs incurred in EP alone, while largely neglecting the other communication costs arising from MP and ESP.

In this paper, we propose two dedicated and complementary schedules (say $S_1$ and $S_2$) for placing communication tasks specifically tailored for the MP+EP+ESP scenario. The highlights of the proposed schedules are: 1) they eliminate redundant computations to reduce computation time, 2) they decrease the overall communication volume to reduce communication time, and 3) they enable overlapping between intra-node and inter-node communications to optimize bandwidth utilization. However, $S_1$ and $S_2$ have distinct communication volumes, which depend on the model and environment configurations, making them suitable for different scenarios. Hence, we develop an automatic and accurate solution to determine which schedule applies during training. To achieve this, we conduct thorough theoretical analyses for the time performance of the baseline schedule, $S_1$, and $S_2$. Based on these analyses, we derive a closed-form solution for choosing $S_1$ or $S_2$ to achieve shorter training time. Integrating the two designed schedules and the automatic schedule decision scheme, we develop a prototype system named Parm atop PyTorch for training large MoE models. 

We perform extensive experiments on an 8-GPU server and a 32-GPU cluster to compare Parm with DeepSpeed-MoE~\cite{DBLP:Rajbhandari22deepspeedmoe}, the current state-of-the-art MoE training system. The experimental results show that our Parm outperforms DeepSpeed-MoE significantly. Specifically, when evaluated on the configured 1296 valid runnable MoE models, Parm achieves 1.13$\times$ to 5.77$\times$ speedups over DeepSpeed-MoE on our testbeds. For two real-world MoE models based on BERT and GPT-2, Parm trains the models 2.98$\times$-3.15$\times$ faster than DeepSpeed-MoE.

% The rest of the paper is organized as follows. We introduce some background and preliminaries in Section~\ref{sec:background}, followed by our proposed two new dedicated schedules in Section~\ref{sec:methods}. We provide thorough performance analyses of our proposed schedules and their trade-off in Section~\ref{sec:formulation}, followed by our developed automatic selection solution in Section~\ref{sec:solution}. Then we demonstrate the evaluation in Section~\ref{sec:evaluation}. Some related works are introduced in Section~\ref{sec:relatedwork}. Finally, we conclude the paper in Section~\ref{sec:conclusion}.

\section{Background and Preliminaries}\label{sec:background}
% This section introduces some background and preliminaries about MoE and distributed training. 
For ease of presentation, we summarize the key notations used throughout the paper in Table~\ref{tab:notations}.

\begin{table}[!ht]
	\centering
	\caption{Notations.}
	\label{tab:notations}
	\begin{tabular}{|l|l|}
		\hline
		Name &  Description \\\cline{1-2}
		\hline
		\hline
		$P$ & \# of workers (or GPUs) in the cluster \\
		$B$ & \# of samples per GPU (or local mini-batch size) \\
		$L$ & \# of tokens per sample (or sequence length) \\
		$T$ & \# of assigned tokens per expert \\
		$E$ & total number of experts \\
		$M$ & embedding size of a token \\
		$H$ & hidden size of the feed-forward layer in experts \\
		$k$ & top-$k$ experts should be selected for each token \\
            $f$ & factor to control expert's maximum token count \\\hline
		% RS & The short for Reduce-Scatter.\\ 
		% AG& The short for AllGather. \\
		% AR & The short for AllReduce. \\
		% A2A & The short for AlltoAll. \\\hline
		% MP & The short for Model-Parallel group. \\
		% DP & The short for Data-Parallel group. \\
		% EP & The short for Expert-Parallel group. \\
		% ES & The short for Expert-Sharding-Parallel group. \\\hline
	\end{tabular}
 \vspace{-2mm}
\end{table}

\begin{figure*}[!t]
	\centering
\includegraphics[width=0.85\linewidth]{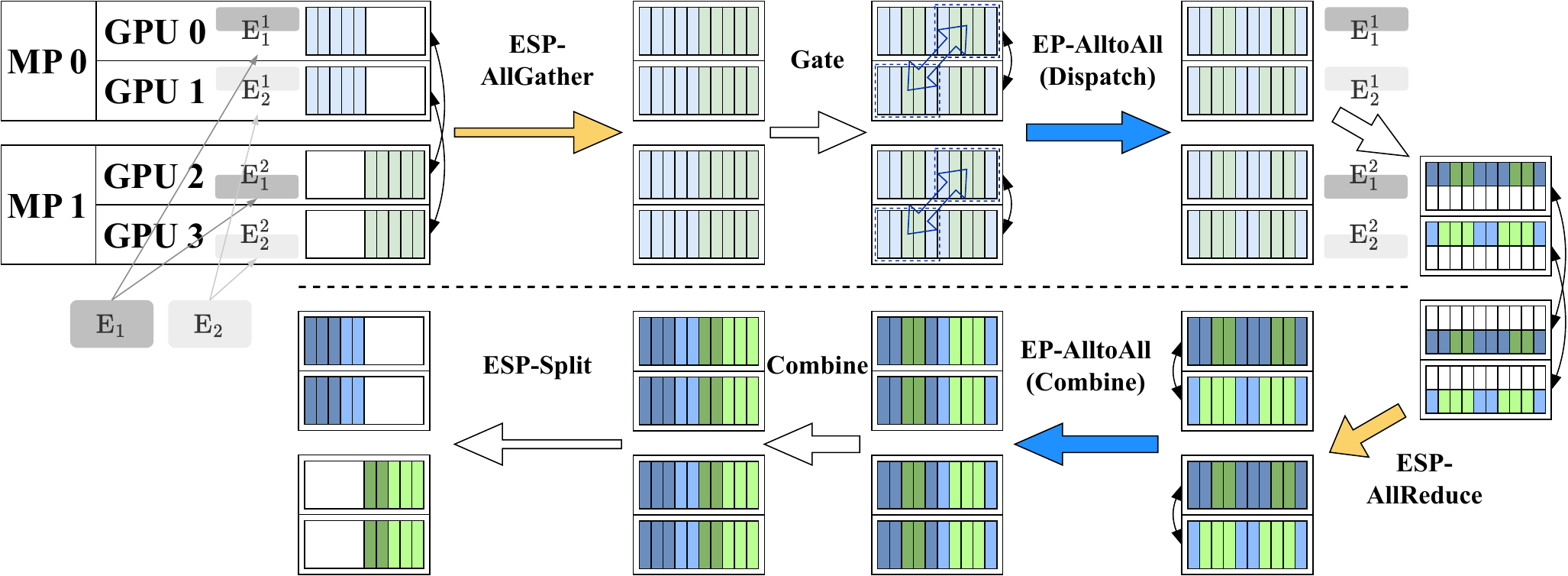}
	\caption{An example of $N_{\text{MP}}=N_{\text{EP}}=N_{\text{ESP}}=2$. The two experts ($\text{E}_1$ and $\text{E}_2$) are distributed to the two EP groups in EP, and each expert is further partitioned into two shards across the ESP group, that is $\text{E}_1\rightarrow[\text{E}_1^1$,$\text{E}_1^2$] are distributed to GPU 0 and GPU2 respectively and $\text{E}_2\rightarrow[\text{E}_2^1$,$\text{E}_2^2$]) are distributed to GPU 2 and GPU 4 respectively. The blue and green rectangles indicate the data tensors and the partially colored part represents a partial sum.}
 \vspace{-2mm}
 \label{fig:full_example}
\end{figure*}

\subsection{Mixture-of-Experts Layer}
Mixture-of-experts (MoE) layers offer a cost-effective approach by selectively activating specific experts for particular inputs, resulting in a sub-linear increase in computational costs when adding more parameters.

Specifically, the MoE layer consists of two primary components: a gating function and a set of $E$ experts. MoE models inherit the dense model and use the MoE layer to replace the dense feed-forward layer (FFN). During training, the input tensor $I$ for the MoE layer has a shape of ($B, L, M$), i.e., $I\in \mathbb{R}^{B\times L\times M}$, where $B$ represents the mini-batch size, $L$ represents the sequence length per sample, and $M$ represents the embedding size of a token. 

\textbf{Gating Function}. The gating function determines which tokens should be processed by each expert. It takes the input tensor $I$ and applies a gating function $g$, which typically consists of a small feed-forward neural network (usually with one linear layer) and a routing function (with a softmax layer and top-$k$ selection). The output tensor $G\in \mathbb{R}^{E\times T\times M}$ is generated, where $T$ is the maximum number of tokens assigned to an expert. 
The softmax function's dynamic nature may lead to an imbalanced workload across different GPUs, with some experts processing more tokens than others. To address this, a capacity factor $f$ is used to limit the maximum number of tokens assigned to an expert, that is $T \coloneqq k\times f\times B\times L / E$. Each row of $G$ ($G[i,:,:]$) corresponds to the data that should be dispatched to expert $i$ ($i=1, 2, \cdots, E$) for computation. Formally, $G=g(I)$.

\textbf{Experts}. An MoE layer consists of $E$ experts, and each expert (say expert $i$) only processes the data in the tensor $G[i, :, :]$ at each iteration during training or inference. All experts have an identical structure and can be executed in parallel with different input tokens. Typically, each expert is a small neural network consisting of two FFN layers followed by an activation function~\cite{DBLP:Lepikhin2021gshard}. The FFN layer has a weight matrix of shape ($M, H$), and the second layer has a shape of ($H, M$), where $H$ is the size of the hidden layer. We denote the expert $i$ as $e_i$, and its output as $Q_i = e_i(G[i,:,:])$. To generate the MoE output, the outputs of different experts are combined into a single tensor $[Q_1, Q_2, ..., Q_{E}]$, which can be reshaped as ($B, L, M$) for further processing in subsequent layers.

\subsection{Paradigms of Parallelism}
DP, PP, MP, EP, and ESP are the five popular parallel paradigms in training large sparse models with a large-scale cluster. As our work focuses on optimizing MP+EP+ESP, and the proposed optimizations can also be integrated with DP and PP, we mainly introduce MP, EP, and ESP. 
% \textbf{Data Parallelism.}
% Data parallelism~\cite{DBLP:dean2012large} is a de-facto parallelism in distributed training by dividing the data across multiple shards along the batch dimension to different processors, particularly when the device memory can accommodate loading the entire model parameters.
% Data Parallelism~\cite{DBLP:dean2012large} accelerates distributed training or inference by splitting the data along the batch dimension into several shards, especially when the device memory is able to load model parameters. We use stochastic gradient descent (SGD) update as an example to illustrate Data Parallelism. A global batch of $B\times P$ samples $X_t$ is distributed to $P$ GPUs at iteration $t$, which means $i-\text{th}$ worker processes $B$ samples $X_t^{(i)}$. Specifically, Data Parallelism can be formalized as: 
% \begin{equation}
% 	W_{t+1}=W_{t}-\eta_t \frac{1}{P}\sum_{i=1}^P\nabla \mathcal{L}\left(W_t, X_t^{(i)}\right),
% 	\nonumber
% \end{equation}
% where $W_{t}$ denotes the model parameters and $\eta_t$ denotes the learning rate. A data-parallel group shorted DP is a set of GPUs that work together to perform Data Parallelism.

\textbf{Model Parallelism.}
Model Parallelism (MP)~\cite{DBLP:dean2012large} (also known as Tensor Parallelism~\cite{narayanan2021efficient} or Operator Parallelism~\cite{DBLP:zheng2022alpa}) partitions model parameters across multiple workers to execute computations in parallel. 
For example, consider a model with two linear layers, $W_1$ and $W_2$, and an activation function $f$. The computation, $Y = W_2f(W_1X)$, is parallelized in $P$-way MP by dividing $W_1$'s columns and $W_2$'s rows. This results in $Y = \sum_{p=1}^{P} W_2^{(p)}f(W_1^{(p)}X)$, allowing parallel execution of each segment. The segments are then aggregated across GPUs using AllReduce.

\textbf{Expert Parallelism.}
A standard implementation of MoE suffers from the issue of a shrinking batch~\cite{DBLP:Shazeer2017outrageously}. When the gating function selects $k$ out of $E$ experts for each token, each expert receives only $T = k\times f / E \times B\times L$ tokens, which is usually much smaller than the local token size $B \times L$. To address this issue, expert parallelism (EP)~\cite{DBLP:Shazeer2017outrageously} is employed, as shown in Fig.~\ref{fig:full_example}. With EP, each device retains only a subset of experts. After passing through the gating function, each row of $G$ ($G[i,:,:]$) on each device corresponds to data that should be dispatched to expert $i$ ($i=1, 2, \cdots, E$). Since the experts may reside on different devices, the dispatch operation is implemented using the AlltoAll collective, known as \textit{AlltoAll Dispatch}. Following the AlltoAll Dispatch, the tokens are subsequently fed to their corresponding experts for computation (i.e., FFNs). Afterward, the outputs from all experts are aggregated using another AlltoAll operation, which is referred to as \textit{AlltoAll Combine}, to be processed for its following layer. We use $N_\text{EP}$ to denote the number of GPUs in an EP group.
% An expert-parallel group shorted EP is a set of GPUs that work together to perform Expert Parallelism. 

\textbf{Expert-Sharding Parallelism.} 
% In this paper, $P$ refers to the total number of GPUs within a cluster. If P is greater than the number of experts, E, then Expert Parallelism employs only E workers while the remaining P-E workers remain idle. To further optimize this process, Expert-Sharding Parallelism is introduced, which divides a single expert's workload and executes it across multiple devices in parallel. To accomplish this objective, Expert-Sharding Parallelism introduces three additional operators: ES-AllGather, ES-AllReduce, and Split, as shown in Figure~(\ref{fig:full_example}). To ensure that the input is uniform across all shards, the Expert-Sharing-Parallel group employs an AllGather primitive. Subsequently, an AllReduce operation is used among the Expert-Sharing-Parallel group to synchronize the output of the expert sharding. Finally, the combined output is split to switch off Expert-Sharding Parallelism. We denote $N_\text{ES}$ as the number of GPUs in an expert-sharding-parallel group.
%It is worth mentioned that the extra communication cost of Moe layers is practically nonnegligible, which may even occupy the dominant position of the overall time on dense-GPU or Google TPU clusters. Previous works have made great efforts to conquer the issue by the overlap between the communication and computing task. However, these works barely try to further wring out the gap among communicative tasks themselves (e.g the inter all-to-all and intra all-gather) which is not rare in the cases with several types of parallel and thus is highlighted in our paper.
In training large-scale MoE models, the number of workers $P$ would be possibly larger than $E$. 
In such a scenario, Expert-Sharding Parallelism (ESP)~\cite{hwang2023tutel,DBLP:Rajbhandari22deepspeedmoe} can be used to balance the workload of all workers. In ESP, an ESP group partitions the experts uniformly to all GPUs in the group, which is similar to MP. In such a way, all workers in the ESP group can compute the outputs of the expert in parallel. However, for one MoE layer with EP and ESP, it also introduces some additional operators including an AllGather operation, an AllReduce operation, and a Split operation in the ESP group, which are denoted as ESP-AllGather, ESP-AllReduce, and ESP-Split respectively. ESP-AllGather is to ensure the input is uniformly sharded to all workers, ESP-AllReduce is required to aggregate the outputs of the expert shards among the ESP group, and ESP-Split is to split the combined outputs to the original structure of the input in MP+EP+ESP. We use $N_\text{ESP}$ to denote the number of GPUs in an ESP group. An example with a two-node cluster, each of which has two GPUs, that uses two-way EP and two-way ESP using two experts, is shown in Fig.~\ref{fig:full_example}, where $N_\text{MP}=N_\text{EP}=N_\text{ESP}=2$ (details in \S\ref{subsec:baselineschedule}). 

% In this paper, we focus on MP+EP+ESP training and design novel schedules for placing the computation and communication tasks to improve the training performance.
\section{Parm: Dedicated Schedules}\label{sec:methods}
\begin{figure}[!t]
	\centering
	\begin{subfigure}[b]{0.48\textwidth}
		\centering
		\includegraphics[width=\textwidth]{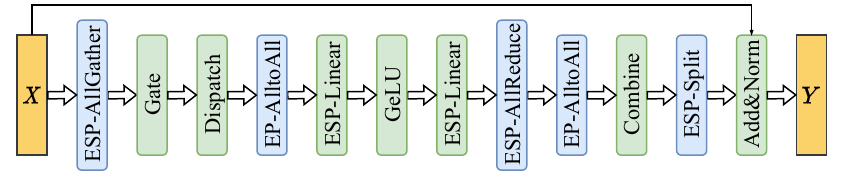}
		\caption{The baseline schedule in existing systems.}
		\label{fig:Baseline}
	\end{subfigure}
 \vspace{4pt}
	
	% \begin{subfigure}[b]{0.8\textwidth}
	% 	\centering
	% 	\includegraphics[width=\textwidth]{figures/a_schedule.pdf}
	% 	\caption{$y=3\sin x$}
	% 	\label{fig:communication-fusion}
	% \end{subfigure}
	
	\begin{subfigure}[b]{0.48\textwidth}
		\centering
		\includegraphics[width=\textwidth]{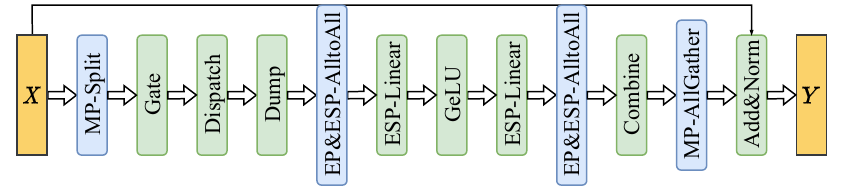}
		\caption{$S_1$: Disable MP before \textit{Gate} and enable it after \textit{Combine}.}
		\label{fig:delete-mp}
	\end{subfigure}
 \vspace{4pt}
	
	\begin{subfigure}[b]{0.48\textwidth}
		\centering
		\includegraphics[width=\textwidth]{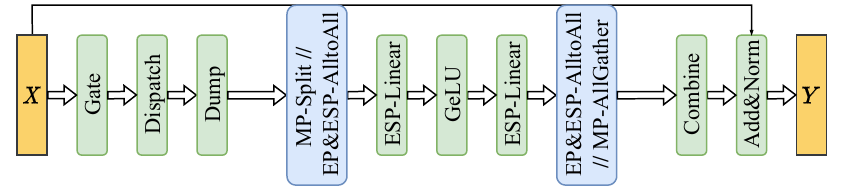}
		\caption{$S_2$: Disable MP after \textit{Gate} and enable MP before \textit{Combine}.}
		\label{fig:overview_overlap}
	\end{subfigure}
	\caption{Three schedules in MP+EP+ESP including (a) the default schedule, (b) our proposed $S_1$ schedule, and (c) our proposed $S_2$ schedule. The yellow color indicates input or output data, the blue color indicates communication operations, and the green color indicates computation operations. Note that the split operations have no communication workload in feed-forward propagation, but they introduce the AllGather communication in backpropagation. That two blue rectangles are overlapped indicates the two operations can be executed in parallel and can be overlapped with each other.}
    \vspace{-2mm}
	\label{fig:schedules}
\end{figure}

\begin{figure}[!t]
	\centering
	\begin{subfigure}[b]{0.48\textwidth}
		\centering
		\includegraphics[width=\textwidth]{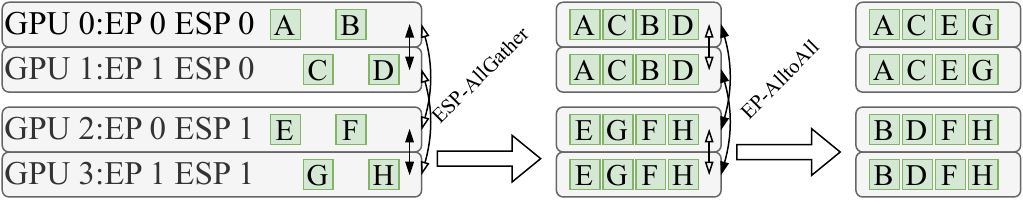}
		\caption{An ESP-AllGather followed by an EP-AlltoAll.}
		\label{fig:observation-1}
	\end{subfigure}
 \vspace{4pt}
 
 \begin{subfigure}[b]{0.48\textwidth}
	\centering
	\includegraphics[width=\textwidth]{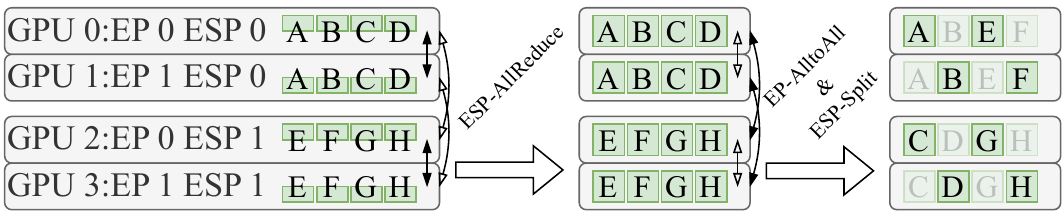}
	\caption{An ESP-AllReduce followed by simultaneous EP-AlltoAll and ESP-Split operations.}
	\label{fig:observation-2}
\end{subfigure}
\vspace{4pt}

	\begin{subfigure}[b]{0.48\textwidth}
		\centering
		\includegraphics[width=\textwidth]{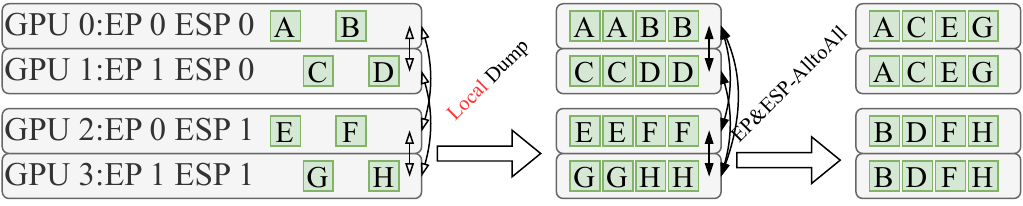}
		\caption{A local data dump followed by an EP\&ESP-AlltoAll.}
		\label{fig:solution-1}
	\end{subfigure}
 \vspace{4pt}
 
\begin{subfigure}[b]{0.48\textwidth}
	\centering
	\includegraphics[width=\textwidth]{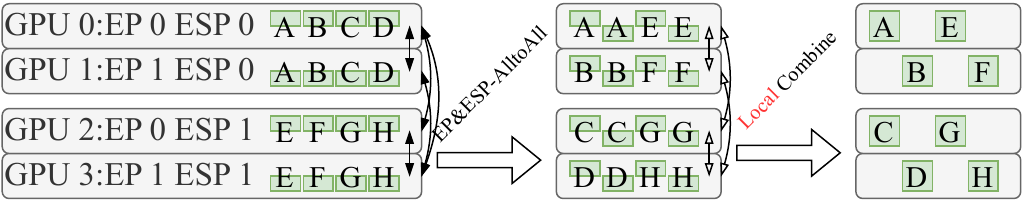}
	\caption{An EP\&ESP-AlltoAll followed by a local data combine.}
	\label{fig:solution-2}
\end{subfigure}
	\caption{Examples of communication patterns with $N_\text{EP}=N_\text{ESP}=2$ under different schedules. The solid arrows indicate requiring communications and the hollow arrow indicates no communications.}
 \vspace{-2mm}
	\label{fig:collectives}
\end{figure}

\begin{figure*}[!ht]
	\centering
		\centering
		\includegraphics[width=0.9\textwidth]{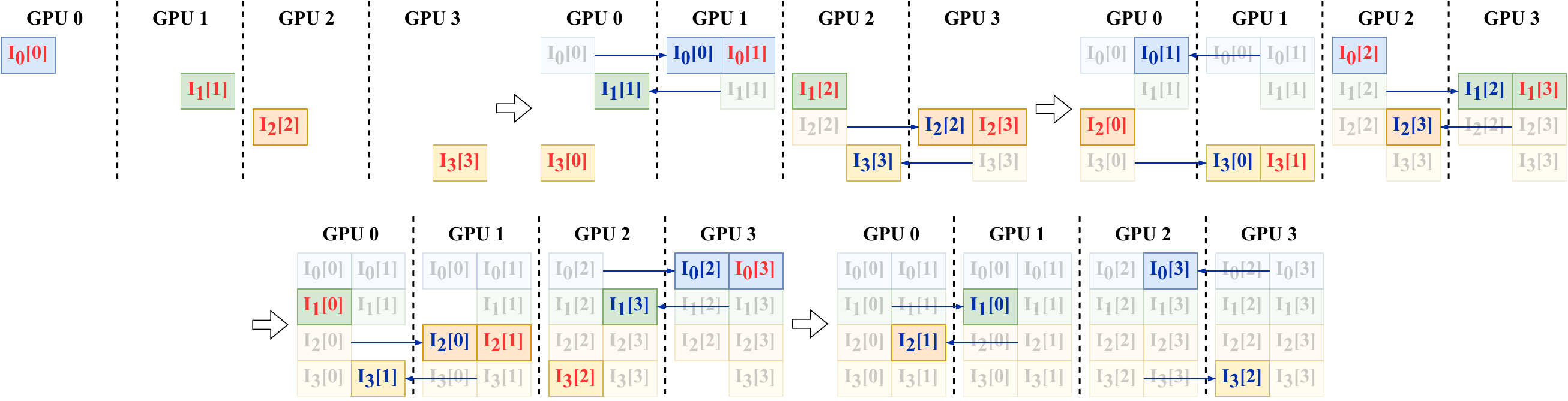}
		\caption{Example of SAA (4-way AlltoAll and 2-way AllGather). Red slices represent data received from AlltoAll at the current turn and blue slices represent the data received from AllGather. Blue arrow represents data transfer using AllGather.}
		\label{fig:overlap-example}
  \vspace{-3mm}
\end{figure*}
% In this section, we begin by discussing our observations on the inefficiencies of the existing schedule. Next, we introduce our two novel schedules. Lastly, we propose a new communication collective that allows for intra-node and inter-node overlap.

\subsection{Observations: Inefficiency of the Baseline Schedule}
In MP+EP+ESP training, an example with $N_\text{MP}=N_\text{EP}=N_\text{ESP}=2$ using the default schedule has been demonstrated in Fig.~\ref{fig:full_example} and it can be generalized as Fig.~\ref{fig:schedules}(a). In the baseline schedule, the input tensor duplication occurs within each MP group, resulting in identical input tokens being processed by the same expert and performing fully identical computations by $N_{\text{MP}}$ times. It means that the default schedule causes three main inefficient operations including two communication operations (ESP-AllGather and ESP-AllReduce in Fig.~\ref{fig:schedules}(a)) and one redundant computing operation (the same expert calculates the same input data by $N_{\text{MP}}$ times). 

\textbf{Observation 1.} As shown in Fig.~\ref{fig:schedules}(a), the baseline schedule first needs to conduct an intra-node ESP-AllGather and followed by an inter-node EP-AlltoAll. Note that the inter-node EP-AlltoAll can only start after the intra-node ESP-AllGather. It means during the communication of each operation, either the intra-node connect or the inter-node connect is idle, making the network resources under-utilized. An example with $N_\text{EP}=N_\text{ESP}=2$ is demonstrated in Fig.~\ref{fig:collectives}(a). 

\textbf{Observation 2.} Similarly, before executing the inter-node EP-AlltoAll operation, an intra-node ESP-AllReduce operation is required for result aggregation. The sequential execution of these two communication operations prevents the overlap of intra-node and inter-node communication, leading to suboptimal utilization. This problem becomes evident when training on GPU clusters without NVLink or NVSwitch. Unfortunately, most researchers do not have access to high-end GPU servers with NVLinks or NVSwitches, making it a common challenge to address. An example of demonstrating how the two operations work is shown in Fig.~\ref{fig:collectives}(b).

Therefore, based on the above observations that cause inefficient communications and duplicate computations, our design goal is to eliminate these three inefficient operations in MP+EP+ESP training.

\subsection{Dedicated Schedules}
Our dedicated schedules revolve around the key idea of temporarily disabling MP (PauseMP) in MoE layers, which brings several benefits: 1) elimination of duplicate computations, 2) significant reduction in communication volume, and 3) replacement of inefficient continuous communications with highly parallel communications, allowing for the overlap of intra-node and inter-node communications. To achieve this, we introduce a Split operation within the MP group (MP-Split) to divide the input tensor of the MoE layer into $N_{\text{MP}}$ parts along the column dimension.
After the MoE layer, we utilize an AllGather operation within the MP group (MP-AllGather) to restore the original input shape. This ensures that each expert processes a unique set of input tokens, thereby optimizing computing resource utilization without redundant operations. In PauseMP, we use a Dump operation to virtually duplicate the data for the expert computation. Thus, the original EP-AlltoAll becomes EP\&ESP-AlltoAll which is a AlltoAll operation (details in \S\ref{subsec:newcollectives}) within both the EP and ESP groups. 

To implement PauseMP in MoE layers, there are two potential schedules. The first schedule involves disabling MP the gate function and enabling it again after combining, as illustrated in Fig.~\ref{fig:schedules}(b). The second schedule entails disabling MP after the gate function and enabling it again before combining as shown in Fig.~\ref{fig:schedules}(c). For clarity, we refer to the above two schedules as $S_1$ and $S_2$, respectively.

\textit{$S_1$ Schedule}: With MP disabled before the gate function, we partition the input data for the gating function, resulting in a reduced size of input data dispatched to experts by $N_{\text{MP}}$ times. Consequently, for the EP\&ESP-AlltoAll collective, the input size becomes $O(BLM\times N_{\text{ESP}} / N_{\text{MP}})$. In comparison to the baseline schedule, $S_1$ effectively reduces both the computation workload and the overall communication volume by $N_{\text{MP}}$ times. While $S_1$ introduces an additional intra-node MP-AllGather, its overhead is significantly smaller than the benefits gained.

% AlltoAll operation in both EP and ESP groups. As shown in Fig.~\ref{fig:schedules}(b), a Dump operation is inserted after Dispatch to virtually duplicate the input data and EP-AlltoAll becomes EP\&ESP-AlltoAll whose input size is $O(BLM)$. After the aggregated results by the Combine operation, an intra-node communication AllGather should be performed in the MP group to gather the data for different GPUs, i.e., MP-AllGather as shown in Fig.~\ref{fig:schedules}(b).

\textit{$S_2$ Schedule}: In the $S_2$ schedule, we move the MP-Split operation before the first EP\&ESP-AlltoAll and place the MP-AllGather after the second EP\&ESP-AlltoAll. Similar to $S_1$, $S_2$ also achieves computation reduction benefits. The communication volume in $S_2$ becomes $O(ETM)$, which is also better than the baseline. Moreover, in $S_2$, the EP\&ESP-AlltoAll operation and the MP-AllGather operation are inter-node communication dominant and intra-node communication dominant operations respectively. This allows us to design a simultaneous AlltoAll and AllGather (SAA) collective to enhance the bandwidth utilization (details in \S\ref{subsec:simultaneous-collective}). 

% ump operation By adjusting the positions of EP\&ESP-AlltoAll from $S_1$, the input size in $S_2$ for the AlltoAll becomes $O(ETM)$. In addition, MP-AllGather is an intra-node communication and EP\&ESP-AlltoAll is a communication operation that mainly consumes inter-node bandwidth. Thus, putting the two collectives together in $S_2$ (as shown in Fig.~\ref{fig:schedules}(c)) can enable the overlap between intra-node data transfer and inter-node data transfer to improve the communication efficiency. It also means MP is disabled after the Dump operation and is enabled after the EP\&ESP-AlltoAll operation. We will design the simultaneous MP-AllGather and EP\&ESP-AlltoAll operations in \S\ref{subsec:simultaneous-collective}. 

The two schedules share identical communication collectives, but they differ in their input dimensions, making them suitable for distinct scenarios. To this end, we devise an automatic strategy which will be elaborated in \S\ref{sec:solution}.

% \textbf{Pausing Model Parallelism} Our study relies on the observation that input tensor duplication occurs within each model-parallel group, resulting in identical input tokens being processed by the same expert and performing fully identical computations $S_{\text{MP}}$ times. To address this issue, we are inspired to temporarily disable model parallelism in MoE layers. We achieve this by using an MP-Split when entering MoE layers and using an MP-AllGather when leaving MoE layers. By doing so, we ensure that each expert processes a unique set of input tokens, which allows for more efficient use of computing resources and faster training times. There are two potential schedules for switching model parallelism on and off. The first option involves turning it off before the gate function and on again after combining. The second option involves turning it off after the gate function and on again before combining. We name these schedules, $S_1$ and $S_2$, respectively, to distinguish between them.

\subsection{EP\&ESP-AlltoAll: One Collective to Replace Two}\label{subsec:newcollectives} 
As shown in Fig.~\ref{fig:collectives}(c)(d), EP\&ESP-AlltoAll performs an AlltoAll operation with both the EP and ESP groups, allowing simultaneous execution of intra-node and inter-node communications. To enable this feature in PauseMP, a local dump operation is necessary to virtually duplicate the data locally before the AlltoAll operation, as shown in Fig.~\ref{fig:collectives}(c). Compared to the sequential execution of ESP-AllGather and EP-AlltoAll in the baseline schedule, EP\&ESP-AlltoAll reduces communication volume and utilizes network resources more efficiently. Similarly, EP\&ESP-AlltoAll also replaces ESP-AllReduce, EP-AlltoAll, and ESP-Split by performing a local combine operation after AlltoAll, as shown in Fig.~\ref{fig:collectives}(d). 
% It generates identical results while enjoying interleaved communication between intra-node and inter-node data transfers. 

% where we visualize the execution of ESP-AllGather and EP-AlltoAll primitives sequentially. We find that the EP-AlltoAll primitive utilizes intra-EP networks while inter-EP networks remain inactive. Similarly, the ESP-AllGather operator exclusively uses intra-ES networks, leaving inter-ES networks idle. According to our observation, we propose a simple yet effective communication fusion. As illustrated in Figure~(\ref{fig:solution-1}), we first duplicate data locally and then use an overall AlltoAll operator in both the expert-parallel group and the expert-sharding-parallel group, named as EP\&ESP-AlltoAll. Compared with executing ESP-AllGather and EP-AlltoAll successively, EP\&ESP-AlltoAll use the network more sufficiently. Similarly, EP\&ESP-AlltoAll also can replace ESP-AllReduce, EP-AlltoAll, and ESP-Split. We first use EP\&ESP-AlltoAll and then combine data locally as shown in Figure~(\ref{fig:solution-2}). The above communication fusion can be used in both of our two schedules. 

\subsection{Simultaneous AlltoAll and AllGather (SAA)}\label{subsec:simultaneous-collective} 
In our $S_2$ schedule, as depicted in Fig.~\ref{fig:schedules}(c), the EP\&ESP-AlltoAll operation can be overlapped with MP-AllGather. The time cost of EP\&ESP-AlltoAll is mainly influenced by the inter-node communication, while MP-AllGather is an intra-node operation that occurs within a single node and is faster in practice. By executing EP\&ESP-AlltoAll and MP-AllGather simultaneously (referred to as SAA), we can better utilize both the intra-node and inter-node communication bandwidths. We implement SAA using NCCL ncclSend and ncclRecv primitives with multiple CUDA streams to enable concurrent communications. An example with a 4-way EP\&ESP-AlltoAll and 2-way MP-AllGather is shown in Fig.\ref{fig:overlap-example}. We divide AlltoAll into four phases, with each phase receiving one data slice from another GPU, indicated in red. The red-marked data slice in one phase becomes the source for the blue-marked AllGather in the next phase. This configuration allows efficient overlapping of data transfers between AlltoAll and AllGather operations, enhancing $S_2$'s performance.

\section{Analysis and Problem Formulation}\label{sec:formulation}
% It is challenging to theoretically analyze the time performance of our proposed communication schedules because it depends on specific physical network topologies. 
To ensure the applicability of our analysis to a broad range of physical network topologies, we adopt minimal and practical assumptions. Specifically, we assume that the cluster consists of multiple homogeneous nodes, each equipped with multiple homogeneous devices. We use $\beta_\text{intra}$ and $\beta_\text{inter}$ to denote two distinct communication speeds for intra-node and inter-node communications, respectively. Without loss of generality, we assume $\beta_\text{intra} > \beta_\text{inter}$. 

\subsection{Time Cost of the Baseline Schedule}\label{subsec:baselineschedule}
The baseline schedule in existing state-of-the-art MoE training systems~\cite{DBLP:Rajbhandari22deepspeedmoe, hwang2023tutel} contains four collective communication operations per MoE layer, as shown in Fig.~\ref{fig:schedules}(a). To enable ESP, the first step is the ESP-AllGather operator, which takes $\text{AG}_{\text{ESP}}\left(BLM\times N_{\text{ESP}}\right)$ time.
Next, the input is dispatched using an EP-AlltoAll primitive, whose time cost is $\text{A2A}_{\text{EP}}\left(ETM\times N_{\text{ESP}}\right)$.
After the expert layers execute their computations, an ESP-AllReduce primitive synchronizes the computation results, and it costs $\text{AR}_{\text{ESP}}\left(ETM\times N_{\text{ESP}}\right)$ time. Finally, another EP-AlltoAll primitive is used to gather the output, which also takes $\text{A2A}_{\text{EP}}\left(ETM\times N_{\text{ESP}}\right)$ time. Note that the ESP-Split operation does not involve communication workload during forward propagation, but it requires an ESP-AllGather operation during backpropagation. Summing it up, the total communication time consumption of the baseline schedule at each training iteration can be formulated as:
\begin{equation}
	\begin{split}
		t_B \coloneqq & \text{AG}_{\text{ESP}}\left(BLM\times N_{\text{ESP}}\right) +\text{AR}_{\text{ESP}}\left(ETM\times N_{\text{ESP}}\right) \\
		&+ 2 \times \text{A2A}_{\text{EP}}\left(ETM\times N_{\text{ESP}}\right).
	\end{split}
 % \vspace{-2mm}
\label{equ:basline}
\end{equation}

\subsection{Time Costs of $S_1$ and $S_2$ Schedules}\label{subsec:time-costs}
Although our proposed Parm schedules are designed for distributed training with MP+EP+ESP, they are also effective in scenarios with EP+ESP alone. In order to demonstrate their effectiveness, we analyze the two schedules separately for the cases when $N_{\text{MP}} = 1$ and $N_{\text{MP}} > 1$, showing that both schedules outperform the baseline.

\noindent\textbf{$N_{\text{MP}} = 1$. }Let's first consider the scenario when $N_{\text{MP}} = 1$, meaning that no MP is used in the distributed training. In this case, pausing MP (i.e., PauseMP) does not directly reduce the communication and computation complexities compared to the baseline schedule. However, it still offers the advantage of enabling communication fusion of ESP-AllGather and EP-AlltoAll into a single communication collective, EP\&ESP-AlltoAll, which is more efficient. Formally, the time cost for dedicated $S_1$ and $S_2$ schedules with EP\&ESP-AlltoAll can be represented as: 
\begin{equation}
t_D \coloneqq 2 \times \text{A2A}_{\text{EP\&ESP}}\left(ETM\times N_{\text{ESP}}\right).
% \vspace{-2mm}
\end{equation}
Our primary objective is to demonstrate the advantages of overlapping ESP-AllGather and EP-AlltoAll, i.e., proving
\begin{equation}
\text{A2A}_{\text{EP\&ESP}}\left(x\right) \leq \text{AG}_{\text{ESP}}\left(x\right) + \text{A2A}_{\text{EP}}\left(x\right),
\label{equ:overlap_a2a}
% \vspace{-2mm}
\end{equation}
where $x$ represents the number of elements in the input tensor. To achieve this, we consider four different cases based on the placements of EP and ESP groups.

\textbf{Case 1: Single Node GPU Configuration.} Here, each GPU exchanges tensors with all other $P$ GPUs, resulting in $(x \times N_{\text{ESP}})$ tensors for EP\&ESP-AlltoAll. Similarly, in the baseline, each GPU communicates with $N_{EP}$ GPUs, also totaling $(x \times N_{\text{ESP}})$ tensors in EP-AlltoAll. Given the equal communication volume $(x \times N_{\text{ESP}})$ and speed $\beta_{\text{intra}}$, both EP-AlltoAll and EP\&ESP-AlltoAll exhibit identical communication times, expressed as:
\begin{equation}
\text{A2A}_{\text{EP\&ESP}}\left(ETM\times N_{\text{ESP}}\right)=\text{A2A}_{\text{EP}}\left(ETM\times N_{\text{ESP}}\right).
% \vspace{-2mm}
\end{equation}
This clearly indicates Equation~(\ref{equ:overlap_a2a}) is satisfied for this case.

\textbf{Case 2: Place ESP groups in the same node.} If all GPUs within the same ESP group are located on the same node, the performance bottleneck of the EP-AlltoAll primitive in the baseline schedule is that some GPU receive/send tensors of dimension $(x \times N_{\text{ESP}} / N_{\text{EP}})$ from/to another GPUs within the same EP group at a lower speed $\beta_{\text{inter}}$. By contrast, in our $S_1$ or $S_2$, the EP\&ESP-AlltoAll operation can be divided into two components: intra-node ESP communication and inter-node ESP communication. The former exhibits the same communication workload and speed as ESP-AllGather. The latter requires each GPU to receive/send tensors with dimensions of $(x / N_{\text{EP}})$ from/to all other GPUs in the other ESP groups, where each group has $\left(N_{\text{EP}} - 1\right) \times N_{\text{ESP}}$ GPUs. Compared to the EP-AlltoAll operation, EP\&ESP-AlltoAll can be viewed as simultaneously processing $N_{\text{ESP}}$ ESP-AlltoAll operations, where each tensor is $N_{\text{ESP}}$ times smaller than that of EP-AlltoAll. Due to the fact that ESP groups are placed in the same node, EP\&ESP-AlltoAll is always faster or equal to EP-AlltoAll, which also makes Equation~(\ref{equ:overlap_a2a}) satisfied.

\textbf{Case 3: Place EP in the same node.} Similarly, if the GPUs within the EP group are located on the same node, the time cost of the ESP-AllGather operation is dominated by the inter-node communication in the baseline schedule, as some GPUs receive/send data (say $x$ dimension) from/to another GPUs within the same ESP group.  
By contrast, in our $S_1$ or $S_2$, the EP\&ESP-AlltoAll operation can be divided into two components: intra-node EP communication and inter-node EP communication. The former exhibits the same communication workload and speed as EP-AlltoAll. The latter entails each GPU receiving/sending tensors with dimensions of $(x / N_{\text{EP}})$  from/to all GPUs in other ESP groups, totaling $\left(N_{\text{ESP}} - 1\right) \times N_{\text{EP}}$ GPUs in total. 
Compared to an ESP-AllGather operator, the latter can be viewed as simultaneously processing $N_{\text{EP}}$ ESP-AllGather, where each tensor is $N_{\text{EP}}$ times smaller than the ESP-AllGather. Due to the fact that expert-parallel groups are placed in the same node, the latter operator is always faster or equal to ESP-AllGather.

\textbf{Case 4: Others.} The time taken by a collective communication depends on the slowest transmission, which is influenced by the uniform workloads among all the participating GPUs in the collective. Given that $\beta_\text{intra} > \beta_\text{inter}$, it is advantageous to place either each ESP group or each EP group in the same node. This ensures that the slowest transmission time for both ESP-AllGather and EP-AlltoAll operations is minimized. Placing them differently would cause an increased slowest transmission time, making the collective less efficient.

Integrating all the above cases, which contain all scenarios in MP+EP+ESP, we prove that Equation~(\ref{equ:overlap_a2a}) always holds. 
Similarly, by replacing ESP-AllGather with an ESP-ReduceScatter, we also can derive
 \begin{equation}
\text{A2A}_{\text{EP\&ESP}}\left(x\right) \leq \text{RS}_{\text{ESP}}\left(x\right) + \text{A2A}_{\text{EP}}\left(x\right)
% \vspace{-2mm}
\end{equation}
since AllReduce can be implemented by an AllGather followed with a ReduceScatter (RS) operator~\cite{barnett1994interprocessor,rabenseifner2004optimization}. Thus, we have: 
\begin{equation}\label{equ:result-1}
	\begin{split}
		 & t_B - t_D \\
  \coloneqq & \text{AG}_{\text{ESP}}\left(BLM\times N_{\text{ESP}}\right) +\text{AR}_{\text{ESP}}\left(ETM\times N_{\text{ESP}}\right) \\
		& +2\times \text{A2A}_{\text{EP}}\left(ETM\times N_{\text{ESP}}\right) \\
            & -2 \times \text{A2A}_{\text{EP\&ESP}}\left(ETM\times N_{\text{ESP}}\right) \\ 
		= & \text{AG}_{\text{ESP}}\left(BLM\times N_{\text{ESP}}\right) + \text{AG}_{\text{ESP}}\left(ETM\times N_{\text{ESP}}\right) \\
		& + \text{A2A}_{\text{EP}}\left(ETM\times N_{\text{ESP}}\right) \\
		& -  \text{A2A}_{\text{EP\&ESP}}\left(ETM\times N_{\text{ESP}}\right) \\
		& +\text{RS}_{\text{ESP}}\left(ETM\times N_{\text{ESP}}\right) + \text{A2A}_{\text{EP}}\left(ETM\times N_{\text{ESP}}\right) \\
		& -  \text{A2A}_{\text{EP\&ESP}}\left(ETM\times N_{\text{ESP}}\right) \\  
			\geq & \text{AG}_{\text{ESP}}\left(BLM\times N_{\text{ESP}}\right),
	\end{split}
\end{equation}
which concludes that our dedicated schedules ($S_1$ or $S_2$) always run faster than the baseline schedule in the scenario of $N_\text{MP} = 1$.

\noindent\textbf{$N_\text{MP} > 1$}. For $N_\text{MP} > 1$, the communication time of the $S_2$ schedule (denoted as $t_{D2}$) is always smaller than the non-overlapping variant, that is,
\begin{equation}
\begin{split}
t_{D2} \coloneqq & \text{A2A}_{\text{EP\&ESP}}\left(ETM\times N_{\text{ESP}} / N_{\text{MP}}\right) \\ & + \text{Overlap}\left(ETM\times N_{\text{ESP}} / N_{\text{MP}}\right)  \\
\leq &  2 \times \text{A2A}_{\text{EP\&ESP}}\left(ETM\times N_{\text{ESP}} / N_{\text{MP}}\right) \\
& + \text{AG}_{\text{MP}}\left(ETM \right),
\end{split}
\end{equation}
where $\text{Overlap}(\cdot)$ denotes the overlap between EP\&ESP-AlltoAll and MP-AllGather. 
An AllGather operator can be viewed as a special AlltoAll operator, which sends the same data to each GPU. We can obtain:
\begin{equation}
\text{AG}_{\text{MP}}\left(x \right) \leq \text{A2A}_{\text{MP}}\left(x \right).  
% \vspace{-2mm}
\end{equation}
Because model parallel are placed in the same node whenever possible, we can derive:
\begin{equation}
\text{AG}_{\text{MP}}\left(x \right) \leq \text{A2A}_{\text{MP}}\left(x \right)  \leq \text{A2A}_{\text{EP\&ESP}}\left(x \right).
% \vspace{-2mm}
\end{equation}
We can calculate the improvement $t_B - t_{D2}$:
\begin{equation}
	\begin{split}
		& t_B - t_{D2} \\
  \coloneqq &  2 \times \text{A2A}_{\text{EP\&ESP}}\left(ETM\times N_{\text{ESP}}\right) \\
		&-2 \times \text{A2A}_{\text{EP\&ESP}}\left(ETM\times N_{\text{ESP}} / N_{\text{MP}} \right) \\
		&-\text{AG}_{\text{MP}}\left(ETM \right)\\
		\ge & 2 \times \text{A2A}_{\text{EP\&ESP}}\left(ETM\times N_{\text{ESP}}\right) \\
		&-2 \times \text{A2A}_{\text{EP\&ESP}}\left(ETM\times N_{\text{ESP}} / N_{\text{MP}} \right) \\
		&-\text{A2A}_{\text{EP\&ESP}}\left(ETM \right)\\
		\ge & \text{A2A}_{\text{EP\&ESP}}\left(2ETMN_{\text{ESP}} \frac{N_{\text{MP}} - 1}{N_{\text{MP}}}\right) \\
		&-\text{A2A}_{\text{EP\&ESP}}\left(ETM\right) \\
		\ge & \text{A2A}_{\text{EP\&ESP}}\left(ETM\right) (\because N_{\text{ESP}} \ge 1, N_{\text{MP}} \ge 2) \\
		&-\text{A2A}_{\text{EP\&ESP}}\left(ETM\right)  \\
            \ge & 0.
	\end{split}
\label{equ:delta}
\end{equation}
Combine Equation~(\ref{equ:result-1}) and Equation~(\ref{equ:delta}), we can conclude that the $S_2$ schedule is always better than the baseline schedule.

When comparing the $S_1$ and $S_2$ schedules, the communication times differ as they depend on the MoE configurations. Specifically, $t_{D1}$ can be formalized as follows:
\begin{equation}
\begin{split}
t_{D1} \coloneqq &  2 \times \text{A2A}_{\text{EP\&ESP}}\left(ETM\times N_{\text{ESP}} / N_{\text{MP}}\right) \\
& + \text{AG}_{\text{MP}}\left(BLM \right).
\end{split}
\end{equation}
For instance, when $N_{\text{ESP}} = 1$ and $T \to 0$, the $S_2$ schedule performs better than $S_1$, as
 $t_{D2} \to 0$ while $t_{D1} \to \text{AG}_{\text{MP}}\left(BLM \right)$. Conversely, when $T \to \infty$, the $S_1$ schedule becomes advantageous because $\text{AG}_{\text{MP}}\left(BLM \right)$ does not increase as $T \to \infty$. This nature of these two schedules leads us to derive how to determine which one should be applied.
\section{Solution}\label{sec:solution}
\subsection{Performance Models}
To automatically choose a better schedule from $S_1$ and $S_2$, we employ the $\alpha$-$\beta$ time performance model~\cite{sarvotham2001connection,DBLP:renggli2019sparcml,he2022fastermoe,shi2023pipemoe}, which represents the communication cost for performing collective communication among a group of workers. In this model, $\alpha$ indicates the startup time to perform the collective operation, and $\beta$ is the transmission time per element. Therefore, we formalize the communication time of collectives by $\alpha + \beta \times x$, where $x$ represents the length of the tensor being transmitted. For example, the time cost of the MP-AllGather operation $\text{AG}_{\text{MP}}(x)$ can be represented as: 
\begin{equation}\label{equ:alltoall-time}
	% \hat{AG}_{\text{MP}}(x) \coloneqq \alpha^{\text{AG}}_{\text{MP}}+ \beta^{\text{AG}}_{\text{MP}}\times x,
	\text{AG}_{\text{MP}}(x) \coloneqq \alpha^{\text{AG}}_{\text{MP}}+ \beta^{\text{AG}}_{\text{MP}}\times x,
% \vspace{-2mm}
\end{equation}
where $\alpha^{\text{AG}}_{\text{MP}}$ and $\beta^{\text{AG}}_{\text{MP}}$ are two value related to the specific network topology in the MP group. To measure $\alpha^{\text{AG}}_{\text{MP}}$ and $\beta^{\text{AG}}_{\text{MP}}$, we can determine the elapsed time using various sizes for the $\text{AG}_{\text{MP}}$ primitives and employ a least square fitting method to estimate them. The collective $\text{A2A}_{\text{EP\&ESP}}$ in different parallel groups has the same form as Equation~(\ref{equ:alltoall-time}) but different $\alpha=\alpha^{\text{A2A}}_{\text{EP\&ESP}}$ and $\beta=\beta^{\text{A2A}}_{\text{EP\&ESP}}$ terms. This performance model primarily addresses homogeneous GPU setups.

\begin{algorithm}[!t]
	\caption{Find the Better Schedule from $S_1$ and $S_2$}\label{algo:solve-schedule}
 \small
	\textbf{Input: }$\alpha^{\text{AG}}_{\text{MP}}$, $\beta^{\text{AG}}_{\text{MP}}$, $\alpha^{\text{A2A}}_{\text{EP\&ESP}}$, $\beta^{\text{A2A}}_{\text{EP\&ESP}}$, $\alpha_{o}$,$\beta_{o}$,$B$,$L$,$E$, $M$,$k$,$f$, $N_\text{ESP}$,  $N_\text{MP}$ \\
    \textbf{Output:} Better schedule from $S_1$ and $S_2$
	\begin{algorithmic}[1]
		\State $x = B \times L \times M $;
		\State $T = k \times f \times B \times L \times M / E$;
		\State $y = E \times T \times M \times N_\text{ESP}$;
		\State ${t}_{D1}=2\times(\alpha^{\text{A2A}}_{\text{EP\&ESP}}+\beta^{\text{A2A}}_{\text{EP\&ESP}} \times y / N_\text{MP} ) + \alpha^{\text{AG}}_{\text{MP}} + \beta^{\text{AG}}_{\text{MP}} \times x$;
		\State ${t}_{D2}=\alpha^{\text{A2A}}_{\text{EP\&ESP}}+\beta^{\text{A2A}}_{\text{EP\&ESP}} \times y / N_\text{MP} + \alpha_{o} + \beta_{o} \times y$;
		\If{${t}_{D1} \leq {t}_{D2}$}
		\State Return $S_1$;
		\Else
		\State Return $S_2$;
		\EndIf
	\end{algorithmic}
\end{algorithm}

\subsection{Automatic Selection Solution}
To derive the automatic selection from $S_1$ and $S_2$, we should quantitatively measure the time difference between $t_{D1}$ and $t_{D2}$. Formally, according to the time performance of each communication collective, $t_{D1}$ can be represented as 
\begin{equation}
\begin{split}
{t}_{D1} \coloneqq &  2 \times {\text{A2A}}_{\text{EP\&ESP}}\left(ETM\times N_{\text{ESP}} / N_{\text{MP}}\right) \\
& + {\text{AG}}_{\text{MP}}\left(BLM \right) \\
= & 2\times\left(\alpha^{\text{A2A}}_{\text{EP\&ESP}}+\beta^{\text{A2A}}_{\text{EP\&ESP}} \times ETM / N_\text{MP} \right) \\
& + \left(\alpha^{\text{AG}}_{\text{MP}} + \beta^{\text{AG}}_{\text{MP}} \times BLM\right).
\end{split}
\end{equation}
Similarly, the time cost of schedule $S_2$ can be represented as
\begin{equation}
\begin{split}
t_{D2} \coloneqq & {\text{A2A}}_{\text{EP\&ESP}}\left(ETM\times N_{\text{ESP}} / N_{\text{MP}}\right) \\
& + {\text{Overlap}}\left(ETM\times N_{\text{ESP}} / N_{\text{MP}}\right) \\
& + {\text{AG}}_{\text{MP}}\left(ETM \right) \\
= & \left(\alpha^{\text{A2A}}_{\text{EP\&ESP}}+\beta^{\text{A2A}}_{\text{EP\&ESP}} \times ETM \times N_\text{ESP} / N_\text{MP} \right) \\
& + \left( \alpha_{o} + \beta_{o} \times ETM \times N_{\text{ESP}} / N_{\text{MP}} \right) \\
& + \left(\alpha^{\text{AG}}_{\text{MP}} + \beta^{\text{AG}}_{\text{MP}} \times ETM\right),
\end{split}
\end{equation}
where $\alpha_{o}$ and $\beta_{o}$ are the $\alpha$ and $\beta$ terms of the overlapped version of EP\&ESP-AlltoAll. Note that the $\alpha$ and $\beta$ terms can be measured in the real-world clusters (\S\ref{subsec:performance-models}). Our strategy is to dynamically compare $t_{D1}$ and $t_{D2}$ online to determine which schedule should be used during training. The algorithm to find the better schedule is shown in Algorithm~\ref{algo:solve-schedule}. 
% After modeling and measuring, we can compare the two schedules and select the optimal schedule based on specific network topologies and hyperparameters of MoE. We use Algorithm~(\ref{algo:solve-schedule}) to summarize our automatic selection solution.

% Compared to our $f\&p$ schedule, the overlapping schedule alters the position of the model parallelism switch, resulting in the replacement of $\text{AG}_{\text{MP}}\left(BLM \right)$ with $\text{AG}_{\text{MP}}\left(ENM \right)$. Based on the theoretical analysis presented above, we can conclude that the overlapping schedule outperforms our $f\&p$ schedule when $f$ is less than one. However, for $f > 1$, it is difficult to definitively determine which schedule is superior without taking into account specific network topologies.

\section{Evaluation}\label{sec:evaluation}

\begin{table}[!t]
	\centering
 \addtolength{\tabcolsep}{-0.5pt}
		\caption{The server configurations in our testbeds.}
		\label{tab:server-config}
		\begin{tabular}{|l|l|l|}
			\hline
			Name & Testbed A & Testbed B \\\hline\hline
			CPU & Dual AMD $\text{EPYC}^{\text{TM}}$ & Dual Intel(R) Xeon(R) \\
                    & 7402 CPU @ 2.80GHz &  Gold 6230 CPU @ 2.10GHz \\\hline
			GPU & 8x Nvidia RTX4090 & 4x Nvidia RTX2080Ti \\
                    & @2.23GHz and 24GB Mem & @1.35GHz and 11GB Mem\\\hline
			Memory & 512GiB DDR4 & 512GB DDR4  \\\hline
			PCIe & 4.0 (x16) & 3.0 (x16)  \\\hline
			Network & - & Mellanox MT27800 Family \\
                        & & (ConnectX-5) @ 100Gb/s \\\hline
		\end{tabular}
\end{table}

\begin{table}[t]
	\centering
	\caption{Configurations of MoE layers.}
	\label{tab:moe-configs}
	\begin{tabular}{|c|c|}
		\hline
		\ & Candidate Values \\\hline\hline
		$P$ & \{8, 16, 32\} \\\hline
		$N_{\text{MP}}$ and $N_{\text{ESP}}$& \{1, 2, 4\} \\\hline
		 % & \{1, 2, 4\} \\\hline
            $B$ & \{2, 4, 8\} \\\hline
		$L$ & \{512, 1024, 2048\} \\\hline
		$H / N_{\text{ES}}$ and $M / N_{\text{ES}}$& \{1024, 2048, 4096\} \\\hline
		 % & \{1024, 2048, 4096\} \\\hline
		$f$ & \{1.2, 2.4\} \\\hline
	\end{tabular}
\vspace{-2mm}
\end{table}

\subsection{Experimental Settings}
\textbf{Testbeds.} We conduct experiments on two testbeds: Testbed A, an 8-GPU node featuring eight Nvidia GeForce RTX4090 GPUs linked through PCIe 4.0x16, and Testbed B, a 32-GPU cluster.
% This cluster consists of eight interconnected nodes, each with four Nvidia GeForce RTX2080Ti GPUs, connected via PCIe 3.0x16 and a 100Gb/s InfiniBand network.
Detailed server configurations are provided in Table~\ref{tab:server-config}.
Due to hardware constraints, notably the absence of NVLink or NVSwitch-equipped GPU servers, our experiments primarily utilized PCIe environments.
%The software environments on both testbed A and B are Ubuntu-18.04 and CUDA-11.3. On testbed A, PyTorch-2.0 and NCCL-2.18 are installed and on testbed B, PyTorch-1.12 and NCCL-2.12 are used.
We compare our Parm with the baseline schedule in DeepSpeed-MoE\footnote{\url{https://github.com/microsoft/DeepSpeed} at version v0.8.1.}.

\textbf{MoE configurations.} Similar to \cite{hwang2023tutel,shi2023pipemoe}, we choose a combination of input parameters whose ranges are shown in Table~\ref{tab:moe-configs} to cover a variety of typical configurations of MoE layers. Note that some cases that require memory larger than the capacity of GPU memory cannot run on our testbeds are excluded, so we runs totally 1296 valid cases. The time for the allreduce of gradients is excluded from the training time, as this part does not benefit from improvements under Amdahl's Law.

\subsection{Performance Models}\label{subsec:performance-models}
\begin{figure}[!ht]
	\centering
	\includegraphics[width=0.7\linewidth]{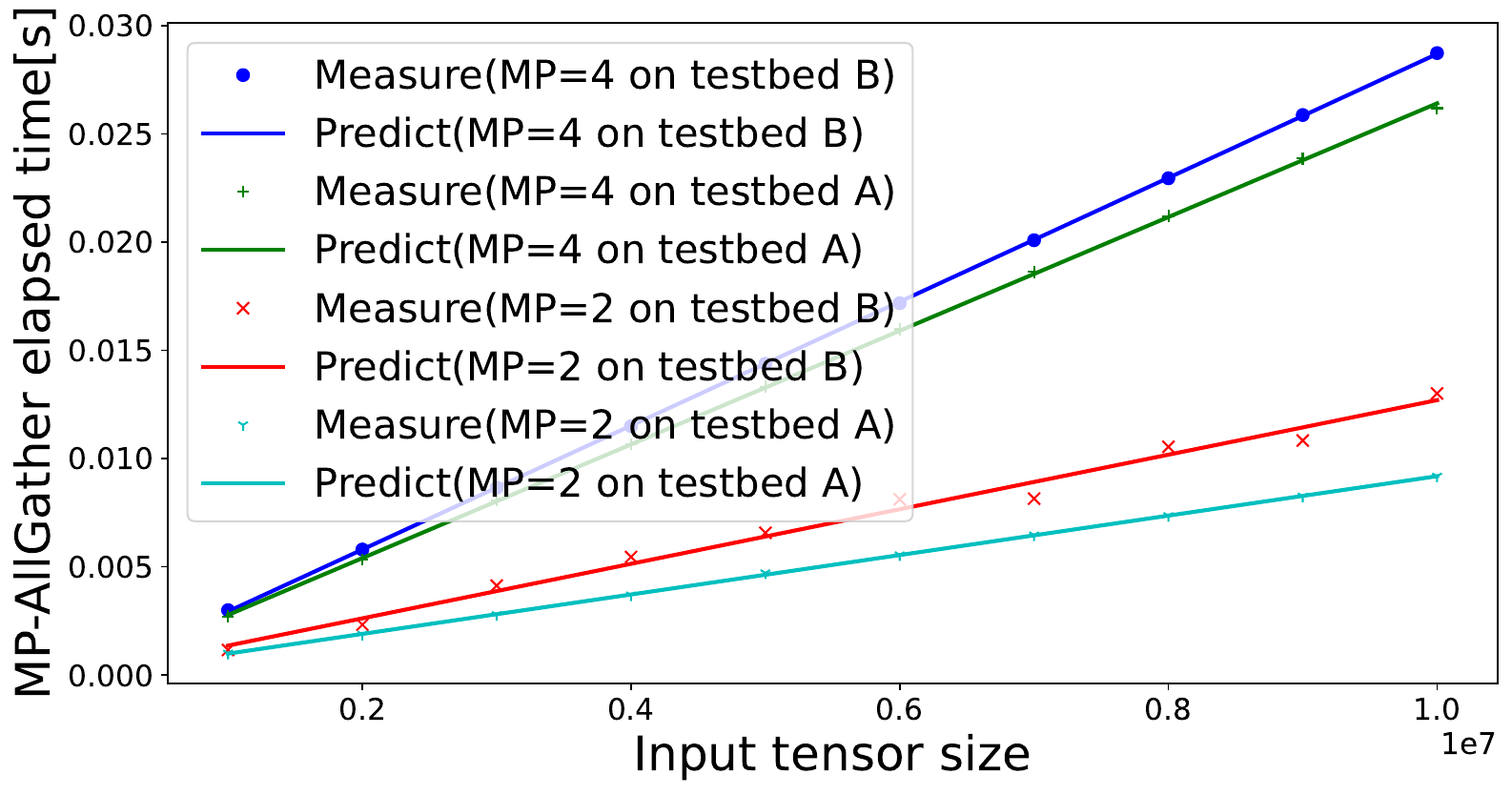}
	\caption{Time performance of collectives on two testbeds.}
\vspace{-3mm}
\label{fig:allgather_time}
\end{figure}
To develop performance models for collective communications, we measure their elapsed time over various message sizes in different parallel groups and collectives. Some partial results are shown in Fig.~\ref{fig:allgather_time}, which shows that the linear model well fits the time cost of collectives and $\alpha_{\text{MP}}^{\text{AG}}=6.64\times10^{-4}
$ and $\beta_{\text{MP}}^{\text{AG}}=5.38\times10^{-10}$ on the 8-GPU server and $\alpha_{\text{MP}}^{\text{AG}}=1.09\times10^{-4}$ and $\beta_{\text{MP}}^{\text{AG}}=7.14\times10^{-10}
$ on the 32-GPU cluster. We use these parameters for Parm in running Algorithm~\ref{algo:solve-schedule} to determine the best schedule.

\subsection{Performance Comparison on Configured MoE Layers}
\begin{table}[!ht]
	\centering
 \addtolength{\tabcolsep}{-0.5pt}
	\caption{Averaged speedups of our schedules over the baseline schedule on testbed A (T-A) and testbed B (T-B) with configured MoE layers in Table~\ref{tab:moe-configs}. For each MoE layer, we measure the average iteration time of 30 iterations and 10 warmup iterations.}
	\label{tab:all-moelayer-experiment}
\begin{tabular}{|cll|c|c|c|c|}
\hline
\multicolumn{3}{|c|}{\multirow{2}{*}{Schedule}} & \multirow{2}{*}{$N_\text{MP}$} & \multirow{2}{*}{$N_\text{ESP}$} & Speedup & Speedup \\ 
\multicolumn{3}{|c|}{} &  &  & (T-A) & (T-B: 8-GPU//16-GPU//32-GPU)\\ \hline\hline
\multicolumn{3}{|c|}{\multirow{4}{*}{$S_1$}} & \multirow{2}{*}{2} & 2 & $2.10\times$ & $2.62\times//2.46\times//2.72\times$ \\ \cline{5-7} 
\multicolumn{3}{|c|}{} &  & 4 & $2.24\times$ & $3.37\times//2.85\times//2.85\times$ \\ \cline{4-7} 
\multicolumn{3}{|c|}{} & \multirow{2}{*}{4} & 2 & $3.72\times$ & $4.18\times//4.12\times//3.92\times$ \\ \cline{5-7} 
\multicolumn{3}{|c|}{} &  & 4 & $4.19\times$ & $5.77\times//5.08\times//4.57\times$ \\ \hline
\multicolumn{3}{|c|}{\multirow{4}{*}{$S_2$}} & \multirow{2}{*}{2} & 2 & $1.99\times$ & $2.42\times//2.20\times//2.57\times$ \\ \cline{5-7} 
\multicolumn{3}{|c|}{} &  & 4 & $2.41\times$ & $3.05\times//2.44\times//2.75\times$ \\ \cline{4-7} 
\multicolumn{3}{|c|}{} & \multirow{2}{*}{4} & 2 & $3.21\times$ & $3.58\times//3.81\times//4.10\times$ \\ \cline{5-7} 
\multicolumn{3}{|c|}{} &  & 4 & $4.12\times$ & $5.38\times//4.82\times//4.91\times$ \\ \hline
\multicolumn{3}{|c|}{\multirow{4}{*}{Parm}} & \multirow{2}{*}{2} & 2 & $2.10\times$ & $2.62\times//2.46\times//2.57\times$ \\ \cline{5-7} 
\multicolumn{3}{|c|}{} &  & 4 & $2.41\times$ & $3.37\times//2.85\times//2.85\times$ \\ \cline{4-7} 
\multicolumn{3}{|c|}{} & \multirow{2}{*}{4} & 2 & $3.70\times$ & $4.18\times//4.12\times//4.06\times$ \\ \cline{5-7} 
\multicolumn{3}{|c|}{} &  & 4 & $4.20\times$ & $5.77\times//5.08\times//4.91\times$ \\ \hline
\end{tabular}
\end{table}
We compare our Parm with DeepSpeed-MoE under a wide range of configurations based on Table~\ref{tab:moe-configs}. Experimental results show that $S_1$, $S_2$, and Parm outperform the baseline schedule in all tested cases, which verifies our theoretical analysis in \S\ref{subsec:time-costs}. The average speedups of our schedules over the baseline schedule are shown in Table~\ref{tab:all-moelayer-experiment}.

\textbf{$S_1$ vs. baseline.} Comparing $S_1$ with the baseline schedule, it is seen that $S_1$ runs 2.1$\times$ to 4.19$\times$ faster on testbed A, and 2.46$\times$ to 5.77$\times$ faster on testbed B. With an increased $N_{\text{MP}}$ or $N_{\text{ESP}}$, the improvement of $S_1$ over the baseline becomes larger. It is mainly because larger $N_{\text{MP}}$ or $N_{\text{ESP}}$ allows $S_1$ to eliminate more duplicate computations and to overlap more communications.

\textbf{$S_2$ vs. baseline.} Similarly, $S_2$ runs 1.99$\times$ to 4.12$\times$ faster on testbed A, and 2.2$\times$ to 5.38$\times$ faster on testbed B than the baseline. It is seen that $S_1$ and $S_2$ may be better than each other under different scenarios. For example, under the setting of $N_{\text{MP}}=4$ and $N_{\text{ESP}}=2$, $S_1$ achieves 3.72$\times$ faster than the baseline, which is larger than the improvement of $S_2$ over the baseline (3.21$\times$) on testbed A. However, on testbed B, the improvement of $S_1$ is smaller than $S_2$ under the same configuration. Note that $S_2$ is also accelerated with SAA (\S\ref{subsec:simultaneous-collective}). Compared with executing AlltoAll and AllGather sequentially (AAS), SAA achieves 1.09\% and 1.12\% improvements over SAA on our two testbeds, respectively.

\textbf{Parm vs. baseline.} As Parm is expected to choose the better one from $S_1$ and $S_2$, Parm achieves higher average speedups over the baseline than $S_1$ and $S_2$, which are 2.1$\times$ to 4.2$\times$ on testbed A and 2.46$\times$ to 5.77$\times$ on testbed B. Since Parm selects $S_1$ or $S_2$ automatically, Parm achieves 8\% and 10.5\% improvements over the worse case of $S_1$ and $S_2$ on testbed A and B, respectively.

\begin{figure}[!ht]
	\centering
	\includegraphics[width=0.6\linewidth]{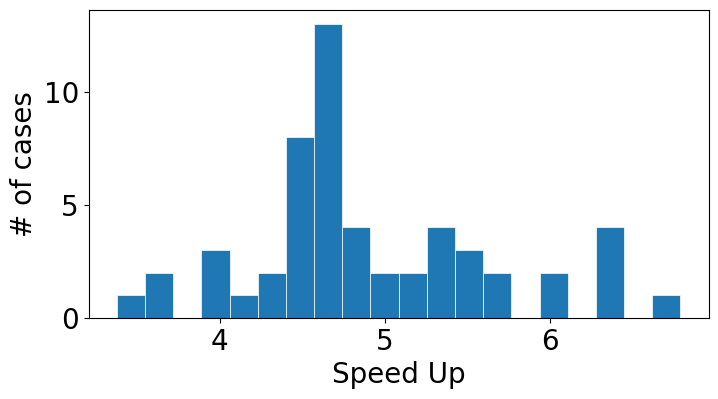}
	\caption{Speedup statistics of Parm over DeepSpeed-MoE on 32 GPUs under $N_{\text{MP}} = 4$ and $N_{\text{ESP}} = 4$.}
	\label{fig:speedup_statistics}
\vspace{-3mm}
\end{figure}

\textbf{Statistics on a cluster. }We also demonstrate the speedup statistics of Parm over DeepSpeed-MoE under the configured MoE layers on our 32-GPU cluster when $N_{\text{MP}} = 4$ and $N_{\text{ESP}} = 4$ as shown in Fig.~\ref{fig:speedup_statistics}. The results show that Parm can outperform DeepSpeed-MoE by 4.91$\times$ on average and the speedup is larger than 4$\times$ in around 89\% cases.

% \textbf{Effects of SAA in $S_2$}. In $S_2$

\subsection{Training Time on Real-world MoE Models}
\begin{table}[!ht]
	\centering
	\caption{Time performance (average iteration time in milliseconds) and the speedup of Parm over DeepSpeed-MoE on real-world MoE models based on BERT-Base and GPT-2.}
	\label{tab:real-world}
\begin{tabular}{|c|c|cl|cl|c|}
\hline
Base Model & Testbed & \multicolumn{2}{c|}{DeepSpeed-MoE} & \multicolumn{2}{c|}{Parm} & Speedup \\ \hline\hline
\multirow{2}{*}{BERT-Base} & A & \multicolumn{2}{c|}{$1733\pm4$} & \multicolumn{2}{c|}{$567\pm2$} & $3.06\times$ \\ \cline{2-7} 
 & B & \multicolumn{2}{c|}{$1920\pm6$} & \multicolumn{2}{c|}{$645\pm3$} & $2.98\times$ \\ \hline
\multirow{2}{*}{GPT-2} & A & \multicolumn{2}{c|}{$1790\pm4$} & \multicolumn{2}{c|}{$581\pm3$} & $3.08\times$ \\ \cline{2-7} 
 & B & \multicolumn{2}{c|}{$2187\pm47$} & \multicolumn{2}{c|}{$695\pm7$} & $3.15\times$ \\ \hline
\end{tabular}
\end{table}
To test Parm's performance in real-world MoE models, we conducted experiments with MoE versions of BERT-Base~\cite{kenton2019bert} and GPT-2~\cite{radford2019language},  using $N_{\text{MP}} = N_{\text{ESP}}= 4$. The number of experts is set to 2 and 8 on testbed A and B respectively. The average iteration time of training is shown in Table~\ref{tab:real-world}, which indicates that Parm outperforms DeepSpeed-MoE by around 3 times.

\section{Related Work}\label{sec:relatedwork}
There are three main orthogonal directions in optimizing the training performance of MoE models: 1) MoE algorithms to choose experts for input tokens to balance workloads for different GPUs~\cite{DBLP:Shazeer2017outrageously,DBLP:Lepikhin2021gshard,lewis2021base,zhou2022mixture,zeng2023scomoe,DBLP:ren2023pangu} and gating functions for improving the model generalization~\cite{zuo2022taming,liu2022gating,li2023the}. 2) AlltoAll algorithms to improve data dispatch and combine efficiency~\cite{all2allnccl2022,DBLP:nie2022hetumoe,DBLP:Rajbhandari22deepspeedmoe,hwang2023tutel,shen2022se,aminabadi2022deepspeed}. 3) Scheduling algorithms to overlap communications with computations~\cite{hwang2023tutel,he2022fastermoe,shi2023pipemoe,li2023lina,zhai2023smartmoe}. Our work mainly focuses on task scheduling, so we mainly introduce the related studies in the third direction.
% The first direction is algorithmic methods that may have various generalization capabilities and the second one is a fundamental optimization for AlltoAll efficiency. The third direction is most related to ours, so we mainly introduce the related studies in this direction.

Tutel~\cite{hwang2023tutel} is a dedicated optimized system for training MoE models with different aspects of optimizations including task scheduling. For pipelining MoE layers, Tutel supports a manually set degree of pipelining or a heuristic search under limited searching space, which may be sub-optimal. FasterMoE~\cite{he2022fastermoe} only allows to partition the input tokens into two groups for pipelining expert computations and AlltoAll communications, which cannot support finer-grain scheduling of these tasks. PipeMoE~\cite{shi2023pipemoe} proposes an optimal partition of input tokens with performance modeling of expert computation and AlltoAll communication. 
SmartMoE~\cite{zhai2023smartmoe} proposes to use a hybrid optimization (i.e., static pools offline and choices to pick within a pool online) by exploring the space of hybrid parallelism with awareness of heterogeneous workloads. Lina~\cite{li2023lina} also studies the communication scheduling and it tries to avoid the network contention of AllReduce and AlltoAll during backpropagation. DeepSpeed-TED~\cite{singh2023hybrid} also reduces data movement but is limited to cases where ESP equals MP. In summary, existing MoE training systems mainly consider the scenario with only DP and EP and optimize its efficiency. They may not be suitable to existing MoE-based large model training as it typically requires to enable multiple dimensions of parallelism including MP, EP, and ESP. 
% We propose to address the communication problem in MP+EP+ESP.
\section{Conclusion}\label{sec:conclusion}
In this paper, we first proposed two dedicated schedules (say $S_1$ and $S_2$) for improving the communication efficiency in distributed training of MoE models under commonly used parallelisms (MP+EP+ESP). The two dedicated schedules not only eliminate redundant computations and communications, but also enable the overlapping between intra-node and inter-node communications. We then provided theoretical analyses for the two schedules and concluded that they are not mutually exclusive. To this end, we further derived an efficient solution to adaptively choose the smaller one from $S_1$ and $S_2$. Integrating the two designed schedules and the automatic schedule decision scheme, we develop a prototype system named Parm atop PyTorch for training large MoE models. We conduct extensive experiments on two testbeds (an 8-GPU server and a 32-GPU cluster) to compare Parm with the state-of-the-art MoE training system DeepSpeed-MoE. Experimental results show that our Parm outperforms DeepSpeed-MoE by 1.13$\times$ to 5.77$\times$ on 1296 configured MoE layers and by 2.98$\times$ to 3.15$\times$ two real-world MoE models based on BERT and GPT-2. The code of Parm can be found at \url{https://github.com/Fragile-azalea/Parm}.

\section*{Acknowledgments}
The research was supported in part by National Natural Science Foundation of China (NSFC) grants under Grant No. 62272122 and 62302123, Guangdong Provincial Key Laboratory of Novel Security Intelligence Technologies under Grant 2022B1212010005, a Hong Kong RIF grant under Grant No. R6021-20, Hong Kong RGC CRF grants under the contracts C2004-21G, C7004-22G, and C1029-22G, and Hong Kong RGC GRF grants under the contracts 16209120, 16200221, and 16207922.

\bibliography{cites}

\begin{thebibliography}{10}
\providecommand{\url}[1]{#1}
\csname url@samestyle\endcsname
\providecommand{\newblock}{\relax}
\providecommand{\bibinfo}[2]{#2}
\providecommand{\BIBentrySTDinterwordspacing}{\spaceskip=0pt\relax}
\providecommand{\BIBentryALTinterwordstretchfactor}{4}
\providecommand{\BIBentryALTinterwordspacing}{\spaceskip=\fontdimen2\font plus
\BIBentryALTinterwordstretchfactor\fontdimen3\font minus \fontdimen4\font\relax}
\providecommand{\BIBforeignlanguage}[2]{{%
\expandafter\ifx\csname l@#1\endcsname\relax
\typeout{** WARNING: IEEEtran.bst: No hyphenation pattern has been}%
\typeout{** loaded for the language `#1'. Using the pattern for}%
\typeout{** the default language instead.}%
\else
\language=\csname l@#1\endcsname
\fi
#2}}
\providecommand{\BIBdecl}{\relax}
\BIBdecl

\bibitem{DBLP:Kaplan2020Scaling}
J.~Kaplan, S.~McCandlish, T.~Henighan, T.~B. Brown, B.~Chess, R.~Child, S.~Gray, A.~Radford, J.~Wu, and D.~Amodei, ``Scaling laws for neural language models,'' \emph{arXiv preprint arXiv:2001.08361}, 2020.

\bibitem{radford2019language}
A.~Radford, J.~Wu, R.~Child, D.~Luan, D.~Amodei, I.~Sutskever \emph{et~al.}, ``Language models are unsupervised multitask learners,'' \emph{OpenAI blog}, vol.~1, no.~8, p.~9, 2019.

\bibitem{DBLP:Brown2020GPT3}
T.~Brown, B.~Mann, N.~Ryder, M.~Subbiah, J.~D. Kaplan, P.~Dhariwal, A.~Neelakantan, P.~Shyam, G.~Sastry, A.~Askell \emph{et~al.}, ``Language models are few-shot learners,'' \emph{Advances in neural information processing systems}, vol.~33, pp. 1877--1901, 2020.

\bibitem{DBLP:Chowdhery2022plam}
A.~Chowdhery, S.~Narang, J.~Devlin, M.~Bosma, G.~Mishra, A.~Roberts, P.~Barham, H.~W. Chung, C.~Sutton, S.~Gehrmann \emph{et~al.}, ``Palm: Scaling language modeling with pathways,'' \emph{Journal of Machine Learning Research}, vol.~24, no. 240, pp. 1--113, 2023.

\bibitem{DBLP:danny2023palme}
D.~Driess, F.~Xia, M.~S. Sajjadi, C.~Lynch, A.~Chowdhery, B.~Ichter, A.~Wahid, J.~Tompson, Q.~Vuong, T.~Yu \emph{et~al.}, ``Palm-e: An embodied multimodal language model,'' \emph{arXiv preprint arXiv:2303.03378}, 2023.

\bibitem{DBLP:Shazeer2017outrageously}
N.~Shazeer, A.~Mirhoseini, K.~Maziarz, A.~Davis, Q.~Le, G.~Hinton, and J.~Dean, ``Outrageously large neural networks: The sparsely-gated mixture-of-experts layer,'' in \emph{International Conference on Learning Representations}, 2016.

\bibitem{DBLP:Lepikhin2021gshard}
D.~Lepikhin, H.~Lee, Y.~Xu, D.~Chen, O.~Firat, Y.~Huang, M.~Krikun, N.~Shazeer, and Z.~Chen, ``Gshard: Scaling giant models with conditional computation and automatic sharding,'' in \emph{International Conference on Learning Representations}, 2020.

\bibitem{DBLP:Fedus2022switch}
W.~Fedus, B.~Zoph, and N.~Shazeer, ``Switch transformers: Scaling to trillion parameter models with simple and efficient sparsity,'' \emph{The Journal of Machine Learning Research}, vol.~23, no.~1, pp. 5232--5270, 2022.

\bibitem{narayanan2021efficient}
D.~Narayanan, M.~Shoeybi, J.~Casper, P.~LeGresley, M.~Patwary, V.~Korthikanti, D.~Vainbrand, P.~Kashinkunti, J.~Bernauer, B.~Catanzaro \emph{et~al.}, ``Efficient large-scale language model training on {GPU} clusters using {Megatron-LM},'' in \emph{Proceedings of the International Conference for High Performance Computing, Networking, Storage and Analysis}, 2021, pp. 1--15.

\bibitem{DBLP:dean2012large}
J.~Dean, G.~Corrado, R.~Monga, K.~Chen, M.~Devin, M.~Mao, M.~Ranzato, A.~Senior, P.~Tucker, K.~Yang \emph{et~al.}, ``Large scale distributed deep networks,'' \emph{Advances in neural information processing systems}, vol.~25, 2012.

\bibitem{DBLP:huang2019gpipe}
Y.~Huang, Y.~Cheng, A.~Bapna, O.~Firat, D.~Chen, M.~Chen, H.~Lee, J.~Ngiam, Q.~V. Le, Y.~Wu \emph{et~al.}, ``Gpipe: Efficient training of giant neural networks using pipeline parallelism,'' \emph{Advances in neural information processing systems}, vol.~32, 2019.

\bibitem{DBLP:Rajbhandari22deepspeedmoe}
S.~Rajbhandari, C.~Li, Z.~Yao, M.~Zhang, R.~Y. Aminabadi, A.~A. Awan, J.~Rasley, and Y.~He, ``Deepspeed-moe: Advancing mixture-of-experts inference and training to power next-generation ai scale,'' in \emph{International Conference on Machine Learning}.\hskip 1em plus 0.5em minus 0.4em\relax PMLR, 2022, pp. 18\,332--18\,346.

\bibitem{hwang2023tutel}
C.~Hwang, W.~Cui, Y.~Xiong, Z.~Yang, Z.~Liu, H.~Hu, Z.~Wang, R.~Salas, J.~Jose, P.~Ram \emph{et~al.}, ``Tutel: Adaptive mixture-of-experts at scale,'' \emph{Proceedings of Machine Learning and Systems}, vol.~5, 2023.

\bibitem{DBLP:liu2022gating}
R.~Liu, Y.~J. Kim, A.~Muzio, and H.~Hassan, ``Gating dropout: Communication-efficient regularization for sparsely activated transformers,'' in \emph{International Conference on Machine Learning}.\hskip 1em plus 0.5em minus 0.4em\relax PMLR, 2022, pp. 13\,782--13\,792.

\bibitem{DBLP:ma2022bagualu}
Z.~Ma, J.~He, J.~Qiu, H.~Cao, Y.~Wang, Z.~Sun, L.~Zheng, H.~Wang, S.~Tang, T.~Zheng \emph{et~al.}, ``Bagualu: targeting brain scale pretrained models with over 37 million cores,'' in \emph{Proceedings of the 27th ACM SIGPLAN Symposium on Principles and Practice of Parallel Programming}, 2022, pp. 192--204.

\bibitem{DBLP:nie2022hetumoe}
X.~Nie, P.~Zhao, X.~Miao, T.~Zhao, and B.~Cui, ``Hetumoe: An efficient trillion-scale mixture-of-expert distributed training system,'' \emph{arXiv preprint arXiv:2203.14685}, 2022.

\bibitem{DBLP:zhang2022accelerating}
S.~Zhang, L.~Diao, C.~Wu, S.~Wang, and W.~Lin, ``Accelerating large-scale distributed neural network training with spmd parallelism,'' in \emph{Proceedings of the 13th Symposium on Cloud Computing}, 2022, pp. 403--418.

\bibitem{shi2023pipemoe}
S.~Shi, X.~Pan, X.~Chu, and B.~Li, ``{PipeMoE}: Accelerating mixture-of-experts through adaptive pipelining,'' in \emph{IEEE INFOCOM 2023-IEEE Conference on Computer Communications}, 2023.

\bibitem{li2023lina}
J.~Li, Y.~Jiang, Y.~Zhu, C.~Wang, and H.~Xu, ``Accelerating distributed $\{$MoE$\}$ training and inference with lina,'' in \emph{USENIX Annual Technical Conference}, 2023, pp. 945--959.

\bibitem{DBLP:zheng2022alpa}
L.~Zheng, Z.~Li, H.~Zhang, Y.~Zhuang, Z.~Chen, Y.~Huang, Y.~Wang, Y.~Xu, D.~Zhuo, E.~P. Xing \emph{et~al.}, ``Alpa: Automating inter-and $\{$Intra-Operator$\}$ parallelism for distributed deep learning,'' in \emph{16th USENIX Symposium on Operating Systems Design and Implementation}, 2022, pp. 559--578.

\bibitem{barnett1994interprocessor}
M.~Barnett, L.~Shuler, R.~van De~Geijn, S.~Gupta, D.~G. Payne, and J.~Watts, ``Interprocessor collective communication library (intercom),'' in \emph{Proceedings of IEEE Scalable High Performance Computing Conference}.\hskip 1em plus 0.5em minus 0.4em\relax IEEE, 1994, pp. 357--364.

\bibitem{rabenseifner2004optimization}
R.~Rabenseifner, ``Optimization of collective reduction operations,'' in \emph{International Conference on Computational Science}.\hskip 1em plus 0.5em minus 0.4em\relax Springer, 2004, pp. 1--9.

\bibitem{sarvotham2001connection}
S.~Sarvotham, R.~Riedi, and R.~Baraniuk, ``Connection-level analysis and modeling of network traffic,'' in \emph{Proceedings of the 1st ACM SIGCOMM Workshop on Internet Measurement}, 2001, pp. 99--103.

\bibitem{DBLP:renggli2019sparcml}
C.~Renggli, S.~Ashkboos, M.~Aghagolzadeh, D.~Alistarh, and T.~Hoefler, ``Sparcml: High-performance sparse communication for machine learning,'' in \emph{Proceedings of the International Conference for High Performance Computing, Networking, Storage and Analysis}, 2019, pp. 1--15.

\bibitem{he2022fastermoe}
J.~He, J.~Zhai, T.~Antunes, H.~Wang, F.~Luo, S.~Shi, and Q.~Li, ``{FasterMoE}: modeling and optimizing training of large-scale dynamic pre-trained models,'' in \emph{Proceedings of the 27th ACM SIGPLAN Symposium on Principles and Practice of Parallel Programming}, 2022, pp. 120--134.

\bibitem{kenton2019bert}
J.~D. M.-W.~C. Kenton and L.~K. Toutanova, ``Bert: Pre-training of deep bidirectional transformers for language understanding,'' in \emph{Proceedings of NAACL-HLT}, 2019, pp. 4171--4186.

\bibitem{lewis2021base}
M.~Lewis, S.~Bhosale, T.~Dettmers, N.~Goyal, and L.~Zettlemoyer, ``{BASE} layers: Simplifying training of large, sparse models,'' in \emph{International Conference on Machine Learning}.\hskip 1em plus 0.5em minus 0.4em\relax PMLR, 2021, pp. 6265--6274.

\bibitem{zhou2022mixture}
Y.~Zhou, T.~Lei, H.~Liu, N.~Du, Y.~Huang, V.~Zhao, A.~Dai, Z.~Chen, Q.~Le, and J.~Laudon, ``Mixture-of-experts with expert choice routing,'' \emph{arXiv preprint arXiv:2202.09368}, 2022.

\bibitem{zeng2023scomoe}
Z.~Zeng and D.~Xiong, ``{SC}omoe: Efficient mixtures of experts with structured communication,'' in \emph{International Conference on Learning Representations}, 2023.

\bibitem{DBLP:ren2023pangu}
X.~Ren, P.~Zhou, X.~Meng, X.~Huang, Y.~Wang, W.~Wang, P.~Li, X.~Zhang, A.~Podolskiy, G.~Arshinov \emph{et~al.}, ``Pangu-{\(\Sigma\)}: Towards trillion parameter language model with sparse heterogeneous computing,'' \emph{arXiv preprint arXiv:2303.10845}, vol.~10, pp. 11--15, 2023.

\bibitem{zuo2022taming}
S.~Zuo, X.~Liu, J.~Jiao, Y.~J. Kim, H.~Hassan, R.~Zhang, J.~Gao, and T.~Zhao, ``Taming sparsely activated transformer with stochastic experts,'' in \emph{International Conference on Learning Representations}, 2022.

\bibitem{liu2022gating}
R.~Liu, Y.~J. Kim, A.~Muzio, and H.~Hassan, ``Gating dropout: Communication-efficient regularization for sparsely activated transformers,'' in \emph{International Conference on Learning Representations}.\hskip 1em plus 0.5em minus 0.4em\relax PMLR, 2022, pp. 13\,782--13\,792.

\bibitem{li2023the}
Z.~Li, C.~You, S.~Bhojanapalli, D.~Li, A.~S. Rawat, S.~J. Reddi, K.~Ye, F.~Chern, F.~Yu, R.~Guo, and S.~Kumar, ``The lazy neuron phenomenon: On emergence of activation sparsity in transformers,'' in \emph{International Conference on Learning Representations}, 2023.

\bibitem{all2allnccl2022}
``Doubling all2all performance with nvidia collective communication library 2.12,'' https://developer.nvidia.com/blog/doubling-all2all-performance-with-nvidia-collective-communication-library-2-12/, accessed: 2022-07-13.

\bibitem{shen2022se}
L.~Shen, Z.~Wu, W.~Gong, H.~Hao, Y.~Bai, H.~Wu, X.~Wu, H.~Xiong, D.~Yu, and Y.~Ma, ``{SE-MoE}: A scalable and efficient mixture-of-experts distributed training and inference system,'' \emph{arXiv preprint arXiv:2205.10034}, 2022.

\bibitem{aminabadi2022deepspeed}
R.~Y. Aminabadi, S.~Rajbhandari, A.~A. Awan, C.~Li, D.~Li, E.~Zheng, O.~Ruwase, S.~Smith, M.~Zhang, J.~Rasley \emph{et~al.}, ``Deepspeed-inference: enabling efficient inference of transformer models at unprecedented scale,'' in \emph{International Conference for High Performance Computing, Networking, Storage and Analysis}.\hskip 1em plus 0.5em minus 0.4em\relax IEEE, 2022, pp. 1--15.

\bibitem{zhai2023smartmoe}
M.~Zhai, J.~He, Z.~Ma, Z.~Zong, R.~Zhang, and J.~Zhai, ``$\{$SmartMoE$\}$: Efficiently training $\{$Sparsely-Activated$\}$ models through combining offline and online parallelization,'' in \emph{USENIX Annual Technical Conference}, 2023, pp. 961--975.

\bibitem{singh2023hybrid}
S.~Singh, O.~Ruwase, A.~A. Awan, S.~Rajbhandari, Y.~He, and A.~Bhatele, ``A hybrid tensor-expert-data parallelism approach to optimize mixture-of-experts training,'' in \emph{Proceedings of the 37th International Conference on Supercomputing}, 2023, pp. 203--214.

\end{thebibliography}
\bibliographystyle{IEEEtran}

\end{document}